\documentclass[useAMS, usenatbib, onecolumn]{mn2e}
\usepackage{amsmath,amssymb}
\usepackage{graphicx,subfigure}
\usepackage{multirow}
\pdfoutput=1

\newcommand{\bb}{\begin{bmatrix}}
\newcommand{\eb}{\end{bmatrix}}

\begin{document}
\pagerange{\pageref{firstpage}--\pageref{lastpage}} \pubyear{2015}

\title[Detection and localization of single-source GWs with PTAs]{Detection and localization of single-source gravitational waves with pulsar timing arrays}
\author[X.-J. Zhu et al.]
{X.-J. Zhu$^{1,2}$\thanks{E-mail: zhuxingjiang@gmail.com},
L. Wen$^{1}$\thanks{E-mail: linqing.wen@uwa.edu.au},
G. Hobbs$^2$,
Y. Zhang$^{3,1,4}$,
Y. Wang$^{5}$,
D. R. Madison$^{6}$,
\newauthor
R. N. Manchester$^2$,
M. Kerr$^{2}$,
P. A. Rosado$^{7}$,
J.-B. Wang$^{8}$
\\
$^1$ School of Physics, University of Western Australia, Crawley WA 6009, Australia\\
$^2$ CSIRO Astronomy and Space Science, PO Box 76, Epping NSW 1710, Australia\\
$^{3}$ Department of Statistics, University of Science and Technology of China, Hefei 230026, China\\
$^{4}$ Department of Statistics, University of Wisconsin-Madison, Madison, WI 53706, USA\\
$^{5}$ School of Physics, Huazhong University of Science and Technology, 1037 Luoyu Road, Wuhan, Hubei 430074, China\\
$^{6}$ Department of Astronomy and Center for Radiophysics and Space Research, Cornell University, Ithaca, NY 14850, USA\\
$^{7}$ Centre for Astrophysics and Supercomputing, Swinburne University of Technology, PO Box 218, Hawthorn Vic 3122, Australia\\
$^{8}$ Xinjiang Astronomical Observatory, CAS, 150 Science 1-Street, Urumqi, Xinjiang 830011, China\\
}
\maketitle \label{firstpage}

\begin{abstract}
Pulsar timing arrays (PTAs) can be used to search for very low frequency ($10^{-9}$--$10^{-7}$ Hz) gravitational waves (GWs). In this paper we present a general method for the detection and localization of single-source GWs using PTAs. We demonstrate the effectiveness of this new method for three types of signals: monochromatic waves as expected from individual supermassive binary black holes in circular orbits, GWs from eccentric binaries and GW bursts. We also test its implementation in realistic data sets that include effects such as uneven sampling and heterogeneous data spans and measurement precision. It is shown that our method, which works in the frequency domain, performs as well as published time-domain methods. In particular, we find it equivalent to the $\mathcal{F}_{e}$-statistic for monochromatic waves. We also discuss the construction of null streams -- data streams that have null response to GWs, and the prospect of using null streams as a consistency check in the case of detected GW signals. Finally, we present sensitivities to individual supermassive binary black holes in eccentric orbits. We find that a monochromatic search that is designed for circular binaries can efficiently detect eccentric binaries with both high and low eccentricities, while a harmonic summing technique provides greater sensitivities only for binaries with moderate eccentricities.
\end{abstract}


\section{Introduction}
The direct detection of gravitational waves (GWs) will have profound implications in physics and astronomy. This may become possible within this decade. In the audio band (10-1000 Hz), second-generation km-scale laser interferometers, such as Advanced LIGO \citep{aLIGO}, Advanced Virgo \citep{aVirgo} and KAGRA \citep{KAGRA}, are about to start scientific observations as early as the second half of 2015 and are expected to detect dozens of compact binary coalescence events per year when they achieve their design sensitivities in around 2020 \citep{lsc_rate10,LVC13localiz}. In the nanohertz frequency range (1-100 nHz), high-precision timing observations of millisecond pulsars (pulsar timing arrays -- PTAs) provide a unique means of detecting GWs. With the concept proposed decades ago \citep{Sazhin1978,Detweiler1979,Hellings_Downs83,Foster_Backer90}, there are now three major PTA projects around the globe, namely, the Parkes Pulsar Timing Array \citep[PPTA;][]{PPTA2013,PPTA13CQG}, the European Pulsar Timing Array \citep[][]{EPTA}, and NANOGrav \citep{NANOGrav}. While these PTAs have individually collected high-quality data spanning $\gtrsim 5$ yrs for $\sim$20 pulsars and produced some astrophysically interesting results \citep[e.g.,][]{PPTA2013Sci}, they have also been combined to form the International Pulsar Timing Array \citep[IPTA;][]{IPTA,IPTAdick13} aiming at significantly enhanced sensitivities.

The primary sources in the PTA frequency band are inspiralling supermassive binary black holes (SMBBHs). It is widely considered that a stochastic GW background due to the combined emission from a large number of individual SMBBHs over cosmological volume \citep[e.g.,][for recent work]{Sesana13GWB,Ravi14GWB} provides the most promising target; indeed, some studies suggested that a detection of this type could occur as early as 2016 \citep{Xavi13CQG}. Analyses of actual PTA data have previously focused on a search for such a GW background, leading to more and more stringent constraints on the fractional energy density of the background \citep{Jenet2006,YardleySGWB,EPTAlimit,NANOGrav2012,PPTA2013Sci}. Over the past few years interest has grown substantially regarding the prospects of detecting single-source GWs using PTAs, for example, for individual SMBBHs \citep{Sesana2009,KJLee2011,Chiara12PRL,Ravi14Single}, for GW memory effects associated with SMBBH mergers \citep{Seto09,CordesJenet12,Dusty14GWM}, for GW bursts \citep{Pitkin12MN} and for unanticipated sources \citep{Cutler14}. In the meantime, many data analysis methods have been proposed in the context of PTAs for single-source GW detection, for example, for monochromatic GWs emitted by SMBBHs in circular orbits \citep{Yardley2010,Babak2012,Ellis2012,EllisBayesian,Taylor14,YanWang14,PPTAcw14}, for GW memory effects \citep{GWM-RvH,GWM_PPTA}, and GW bursts \citep{Finn10ApJ,Deng14PRD}. A few papers have presented searches in real PTA data for GWs from circular SMBBHs and yielded steadily improved upper limits on the GW strain amplitude \citep{Yardley2010,NANOcw14,PPTAcw14}. While circular binaries emit GWs at the second harmonic of the orbital frequency, eccentric binaries radiate at multiple harmonics. \citet{JenetWen04} first derived the expression of pulsar timing signals produced by eccentric binaries and developed a framework in which pulsar timing observations can be used to constrain properties of SMBBHs.

In our previous work  \citep{PPTAcw14}, we performed an all-sky search for GWs from SMBBHs in circular orbits using the latest PPTA data set. Here we adapt the network analysis method used in the context of ground-based interferometers \citep[e.g.,][]{Pai01,WenSchutz05,WenSchutz12,Wen08IJMPD,Klimenko08,Sutton10} as a general method for detection and localization of single-source GWs using PTAs. In particular, we consider the following types of sources: (1) SMBBHs in circular orbits; (2) eccentric SMBBHs; and (3) GW bursts. We demonstrate the effectiveness of this method using synthetic data sets that contain both idealized and realistic observations.

The organization of this paper is as follows. In section \ref{PTA-GWs} we briefly review the features of pulsar timing signals caused by single-source GWs and discuss the signal models used in this work. In section \ref{SVDmethod} we present the mathematical framework of our method and propose practical detection statistics. The relation to the method used in our previous work is also discussed. Using idealized simulations we show examples of detection, localization and waveform estimation in section \ref{Sec-Examp}. We demonstrate the implementation of the method in realistic data sets in section \ref{Sec-Realis}. In section \ref{Sec-SenSMBBH} we present sensitivities to eccentric SMBBHs. In particular we compare two detection strategies towards the detection of eccentric binaries -- a monochromatic search and a harmonic summing technique. Finally, we summarise and outline future work in section \ref{Sec-Conclu}.

\section{Single-source GWs and pulsar timing}
\label{PTA-GWs}
In pulsar timing, the times of arrival (ToAs) of radio pulses from millisecond pulsars are measured and compared with predictions based on a timing model that describes the pulsar's rotation (e.g., spin period and spin-down rate), the relative geometry between the pulsar and the observer, and other effects such as dispersion of radio waves due to interstellar medium \citep{LorimerKramer05,TEMPO2Edwards}. The differences between the measurements and predictions are called timing residuals. These residuals may be partly attributed to effects of GWs as they are not normally included in a timing model. Observations of a single pulsar cannot make a definite detection of a GW, but can lead to constraints on the strength of potential GWs \citep[see, e.g.,][for a recent work]{YiShuXu14MN}. By timing an array of pulsars, GWs can be unambiguously detected by searching for correlated signatures among different pulsars. Typical PTA observations have a sampling interval of weeks and span over $\sim$10 yr, implying a sensitive GW frequency range of $\sim$1--100 nHz.

In this work we are interested in single-source GWs, i.e., those coming from some particular directions in the sky. Timing residuals of such an origin can be generally written as:
\begin{equation}
\label{TR1}
r(t,\hat{\Omega}) = F_{+}(\hat{\Omega})\Delta A_{+}(t) + F_{\times}(\hat{\Omega}) \Delta A_{\times}(t),
\end{equation}
where $\hat{\Omega}$ is a unit vector pointing from the GW source to the observer. The two functions $F_{+}(\hat{\Omega})$ and $F_{\times}(\hat{\Omega})$ are the geometric factors as given by \citep[e.g.,][]{KJLee2011}:
\begin{equation}
F_{+}(\hat{\Omega}) = \frac{1}{4(1-\cos\theta)}\{(1+\sin^2 \delta)\cos^2 \delta_p \cos[2(\alpha-\alpha_p)] - \sin2\delta \sin2\delta_p\cos(\alpha-\alpha_p) + \cos^2 \delta (2-3\cos^2 \delta_p)\}
\label{Fp}
\end{equation}
\begin{equation}
F_{\times}(\hat{\Omega}) = \frac{1}{2(1-\cos\theta)}\{\cos \delta \sin 2\delta_p \sin(\alpha-\alpha_p) - \sin \delta \cos^2 \delta_p \sin[2(\alpha-\alpha_p)]\} ,
\label{Fc}
\end{equation}
where $\cos\theta = \cos\delta \cos\delta_p \cos(\alpha-\alpha_p)+\sin\delta \sin\delta_p$ with $\theta$ being the opening angle between the GW source and pulsar with respect to the observer, $\alpha$ ($\alpha_p$) and $\delta$ ($\delta_p$) are the right ascension and declination of the GW source (pulsar) respectively. Note that we separate the polarization angle from $F_{+}(\hat{\Omega})$ and $F_{\times}(\hat{\Omega})$ and put it into $\Delta A_{+,\times}(t)$ for convenience of later analyses. The two functions $F_{+}(\hat{\Omega})$ and $F_{\times}(\hat{\Omega})$ are analagous to the antenna pattern functions used in the context of laser interferometric GW detection \citep{Thorne87}.

Because the wavelengths of GWs in the PTA band are much smaller than the pulsar-Earth distance, GW induced timing residuals are the combination of two terms -- the Earth term $A_{+,\times}(t)$ and the pulsar term $A_{+,\times}(t_p)$ \citep[see, e.g.,][]{JenetWen04}:
\begin{equation}
\label{TR2}
\Delta A_{+,\times}(t) = A_{+,\times}(t)-A_{+,\times}(t_p)
\end{equation}
\begin{equation}
\label{TpTe}
t_p = t-d_p(1-\cos\theta)/c,
\end{equation}
where $d_p$ is the pulsar distance and we have adopted the plane wave approximation. $A_{+}(t)$ and $A_{\times}(t)$ are source-dependent functions. Throughout this paper, we only consider the correlated Earth-term signals. In fact for the case of continuous waves as expected from SMBBHs, pulsar terms will act as an extra source of uncorrelated noise for different pulsars. For GW bursts, whose duration is smaller than the data span, it is very unlikely that pulsar terms and Earth terms are simultaneously present in timing residuals for one particular pulsar unless the source sky direction is very close to that pulsar. Below we briefly discuss the signal properties for the three types of sources considered in this paper.

At the leading Newtonian order SMBBHs in circular orbits are expected to generate pulsar timing signals with the following forms \citep[see, e.g.,][]{PPTAcw14}:
\begin{equation}
A_{+}(t) = \frac{h_0}{2\pi f} \left[(1+\cos ^{2}\iota) \cos2\psi \sin(2\pi f t+\phi_0)+2\cos\iota \sin2\psi \cos(2\pi f t+\phi_0)\right]
\label{Ap}
\end{equation}
\begin{equation}
A_{\times}(t) = \frac{h_0}{2\pi f} \left[(1+\cos ^{2}\iota) \sin2\psi \sin(2\pi f t+\phi_0)-2\cos\iota \cos2\psi \cos(2\pi f t+\phi_0)\right],
\label{Ac}
\end{equation}
where the GW strain amplitude is $h_0=2 (G M_c)^{5/3} (\pi f)^{2/3} c^{-4} d_L^{-1}$ with $d_L$ being the luminosity distance of the source and $M_c$ being the chirp mass defined as $M_c^{5/3} = m_1 m_2(m_1+m_2)^{-1/3}$ where $m_1$ and $m_2$ are the binary component masses, $f$ is the GW frequency, $\iota$ is the inclination angle of the binary orbit, $\psi$ is the GW polarization angle, $\phi_0$ is a phase constant. Note that we have neglected the frequency evolution over the observation span (typically $\sim 10$ yr). This represents the majority of circular binaries that are observable for PTAs as shown in \citet{Sesana2010}.

Although it is well known that radiation of GWs circularizes the binary orbits \citep{Peters-Mathews63}, the assumption of circular orbits is not always appropriate. Recent models for the SMBBH population including the effects of binary environments on orbital evolution suggest that eccentricity is important for GW frequencies $\lesssim 10^{-8}$ Hz \citep{Sesana13CQG,Ravi14GWB}. In this work we use the expressions of GW-induced timing residuals for eccentric SMBBHs as given in \citet{JenetWen04}. It is interesting to note that eccentric binaries emit GWs at multiple harmonics of the binary orbital frequency. At low eccentricities the emission is dominated by the second harmonic, while for high eccentricities the orbital frequency itself will dominate.

Generally a GW burst is defined as a transient signal with a duration smaller than the observation span. This may be the only information we know about the source. For this reason we use a simple but general sine-Gaussian model for timing residuals induced by GW bursts:
\begin{equation}
A_{+}(t) = A\exp\left(-\frac{(t-t_0)^2}{2\tau^2}\right)\left\{(1+\cos ^{2}\iota)\cos2\psi \cos[2\pi f_{0} (t-t_0)+\phi_0]-2\cos\iota \sin2\psi\sin[2\pi f_{0} (t-t_0)+\phi_0]\right\}
\label{ApGWb}
\end{equation}
\begin{equation}
A_{\times}(t) = A\exp\left(-\frac{(t-t_0)^2}{2\tau^2}\right)\left\{(1+\cos ^{2}\iota)\sin2\psi \cos[2\pi f_{0} (t-t_0)+\phi_0]+2\cos\iota \cos2\psi\sin[2\pi f_{0} (t-t_0)+\phi_0]\right\},
\label{AcGWb}
\end{equation}
where $A$ is the signal amplitude (in seconds), $\tau$ is the Gaussian width, $\iota$ is the source inclination angle, $f_{0}$ is the central frequency, $t_0$ and $\phi_0$ is the time and phase at the midpoint of the burst respectively. The sine-Gaussian model used here can represent qualitatively the signals for parabolic encounters of two massive black holes as studied in \citet{Finn10ApJ} and more recently in \citet{Deng14PRD}. The detection method that we will describe in the next section should apply equally well to other burst sources such as cosmic (super)string cusps and kinks \citep[e.g.,][]{Vilenkin81,Damour00,Siemens06} and triplets of supermassive black holes \citep{TripleBH10}.

\section{The data analysis method}
\label{SVDmethod}
In this section, we describe how the singular value decomposition (SVD) can be used in a general method for the detection and localization of single-source GWs using PTAs. The method is adapted from the coherent network analysis method used in the context of ground-based GW interferometers. There are some important features that are unique to PTA observations, e.g., (1) PTA data are irregularly sampled in contrast to nearly continuous sampling for ground-based experiments; and (2) a least-squares fitting process is performed to obtain estimates of timing parameters such as the pulsar's spin period and its first time derivative, pulsar position and proper motion, etc. As we show later in this section, the SVD method proposed here relies on transforming the timing residuals of each pulsar to the frequency domain. For the idealized simulations (assuming even sampling and white Gaussian noise with equal error bars) used in section \ref{Sec-Examp}, a discrete Fourier Transform was used. In section \ref{Sec-Realis} for more realistic data sets we adopt a maximum-likelihood-based method to estimate Fourier components of the timing residuals making use of the noise covariance matrix \citep[e.g., section 3 in][]{PPTAcw14}. This is equivalent to the least-squares spectral analysis method \citep[see, e.g.,][]{Coles2011}. Our method works with post-fit timing residuals, i.e., after fitting ToAs for timing models of individual pulsars. The effects of the fitting process on our results will be discussed in section \ref{Sec-Realis}.

For a given source direction, the timing residuals from an array of $N_{p}$ pulsars can be generally written in the frequency domain as:
\begin{equation}
\label{data}
\mathbf{d_{k}}=\mathbf{F_{k}}\mathbf{A_{k}}+\mathbf{n_{k}},
\end{equation}
where the index $k$ denotes the $k$-th frequency bin, $\mathbf{d_{k}}$ are timing residual data, $\mathbf{F_{k}}$ is the response matrix, $\mathbf{A_{k}}$ are GW waveforms and $\mathbf{n_{k}}$ is the timing noise. The data are whitened\footnote{Here it is assumed that noise for different pulsars is uncorrelated. If not the full covariance matrix should be used in a whitening process.} so that
\begin{equation}
\label{whitedata}
\mathbf{d_{k}}= \bb  d_{1k}/\sigma_{1k}\\d_{2k}/\sigma_{2k}\\ \vdots \\ d_{N_pk}/\sigma_{N_pk}\eb,
\mathbf{A_k} = \bb A_{+k}\\ A_{\times k}\eb,
\mathbf{n_k}=\bb  n_{1k}/\sigma_{1k}\\n_{2k}/\sigma_{2k}\\ \vdots \\ n_{N_pk}/\sigma_{N_pk}\eb,
\end{equation}
where $\sigma^2_{ik}$ is the noise variance of the $i$-th pulsar at the $k$-th frequency bin. Here $\mathbf{d_{k}}$, $\mathbf{A_k}$ and $\mathbf{n_k}$ are all complex vectors, while the real whitened response matrix $\mathbf{F_{k}}$ is defined as
\begin{equation}
\label{whiteF}
\mathbf{F_{k}} = \bb F^+_1/\sigma_{1k}  & F^\times_1 /\sigma_{1k}  \\ F^+_2/\sigma_{2k}  & F^\times_2/\sigma_{2k} \\ \vdots & \vdots \\ F^+_{N_p}/\sigma_{N_p k}  & F^\times_{N_p} /\sigma_{N_p k} \eb.
\end{equation}
For simplicity we will hereafter suppress the index $k$ while keeping in mind that equations (\ref{data}-\ref{whiteF}) apply to each frequency bin in the analysis. Data for all frequencies can be stacked (e.g., in order of increasing frequency) to preserve the format of equation (\ref{data}), in which case $\mathbf{F}$ is a block-diagonal matrix (with a dimension of $N_pN_k \times 2N_k$ where $N_k$ is the number of frequency bins) of $\mathbf{F_{k}}$.

The SVD of the response matrix $\mathbf{F}$ can be written as
\begin{equation}
\label{svdF}
\mathbf{F}=\mathbf{U}\mathbf{S}\mathbf{V}^{\ast},
\mathbf{S} = \bb s_{1} &  0 \\ 0 & s_{2} \\ 0 & 0 \\ \vdots & \vdots \\ 0 & 0 \eb,
\end{equation}
where $\mathbf{U}$ and $\mathbf{V}$ are unitary matrices with dimensions of $N_p \times N_p$ and $2 \times 2$ respectively, and the symbol $\ast$ denotes the conjugate transpose. Singular values in $\mathbf{S}$ are ranked such that $s_{1}\geqslant s_{2}\geqslant 0$. We can then construct new data streams as follows:
\begin{equation}
\label{new-data}
\mathbf{\tilde{d}}=\mathbf{U}^{\ast}\mathbf{d}, \,\,  \mathbf{\tilde{A}}=\mathbf{V}^{\ast} \mathbf{A}, \,\,  \mathbf{\tilde{n}}=\mathbf{U}^{\ast}\mathbf{n} .
\end{equation}
One can find that $\mathbf{\tilde{d}}= \mathbf{S}\mathbf{\tilde{A}}+\mathbf{\tilde{n}}$, and explicitly
\begin{equation}
\label{new-data1}
\mathbf{\tilde{d}}=\bb  s_{1}\left(\mathbf{V}^{\ast} \mathbf{A}\right)_{1}+\mathbf{\tilde{n}}_{1} \\ s_{2}\left(\mathbf{V}^{\ast} \mathbf{A}\right)_{2}+\mathbf{\tilde{n}}_{2} \\ \mathbf{\tilde{n}}_{3} \\ \vdots \\ \mathbf{\tilde{n}}_{N_{p}}\eb .
\end{equation}
In the absence of GW signals, the real and imaginary parts for each element of $\mathbf{\tilde{d}}$ independently follow the standard Gaussian distribution. Here $\mathbf{\tilde{d}}_1$ and $\mathbf{\tilde{d}}_2$ are referred to the two signal streams\footnote{Here the indices $1$ and $2$ refer to the first two terms in the new data streams $\mathbf{\tilde{d}}$. More indices are used for time and frequency when necessary.} since they contain all information about GW signals if present, while the remaining terms are `null streams' (denoted by $\mathbf{\tilde{d}}_{\rm{null}}$) since they have a null response to GWs. It has been shown that, in the context of GW detection with ground-based laser interferometers, null streams can be used as a consistency check on whether a candidate GW event is produced by detector noise or by a real GW \citep{WenSchutz05} and `semi-null streams' (i.e., $\mathbf{\tilde{d}}_2$ if $s_{1}\gg s_{2}$) can be included to improve the angular
resolution \citep{Wen08IJMPD,WenFanChen08}. In this paper we only consider Earth terms in our signal model so the null streams (as constructed in the current way) are indeed `null' but in reality they would have some response to the pulsar-term signals. In a future study pulsar terms will be incorporated in the detection framework with the addition of $N_p$ free parameters for pulsar distances in the response matrix $\mathbf{F}$.

It is straightforward to show that the maximum likelihood estimator for the GW waveform is
\begin{equation}
\label{Aest}
\hat{\mathbf{A}}=\bar{\mathbf{F}}\mathbf{d},\, {\rm{where}} \,\, \bar{\mathbf{F}}=\mathbf{V}\bar{\mathbf{S}}\mathbf{U}^{\ast}, \,\bar{\mathbf{S}} = \bb 1/s_{1} &  0 & 0 & \hdots & 0 \\ 0 & 1/s_{2} & 0 & \hdots & 0 \eb.
\end{equation}
The matrix $\bar{\mathbf{F}}$ is the Moore-Penrose pseudoinverse of the response matrix $\mathbf{F}$. The covariance matrix for the estimated waveform is
\begin{equation}
\label{Avar}
{\rm{var}}(\hat{\mathbf{A}})= \left(\mathbf{V}\mathbf{S}^{T}\mathbf{S}\mathbf{V}^{\ast}\right)^{-1}.
\end{equation}
It is interesting to note that the statistical uncertainties of the estimated waveforms are a linear combination of $s_{1}^{-2}$ and $s_{2}^{-2}$. Estimation of physical parameters can be obtained based on the estimated waveforms $\hat{\mathbf{A}}$ and we leave this to a future study.

Now we propose our detection statistics for the three types of signals discussed in the previous section. For monochromatic GWs, the detection statistic can be written as:
\begin{equation}
\label{DS-mon}
\mathcal{P}_{\rm{mon}}= |\mathbf{\tilde{d}}_1|^{2}+|\mathbf{\tilde{d}}_2|^{2}.
\end{equation}
It is important to note that: (1) in the absence of GW signals, $\mathcal{P}_{\rm{mon}}$ follows a $\chi^2$ distribution with 4 degrees of freedom; and (2) the detection statistic applies to a given GW frequency and source sky location. In practice when such information is unknown, a search is usually performed to find the maximum statistic.

For signals produced by eccentric binaries, we use a harmonic summing technique for which the detection statistic is given by:
\begin{equation}
\label{DS-ecc}
\mathcal{P}_{\rm{ecc}}= \sum^{N_h}_{j=1} \left(|\mathbf{\tilde{d}}_{1,jk_{0}}|^{2}+|\mathbf{\tilde{d}}_{2,jk_{0}}|^{2}\right),
\end{equation}
where $N_h$ corresponds to the highest harmonics considered in the search, $k_0$ is the bin number for the binary orbital frequency. In the absence of GW signals, $\mathcal{P}_{\rm{ecc}}$ follows a $\chi^2$ distribution with $4N_{h}$ degrees of freedom. In practice $N_h$ should be determined as the one that gives the lowest false alarm probability (FAP). When the orbital frequency is unknown one should search over all possible frequencies to find the maximum statistic.

For GW bursts with unknown waveforms, we adopt a time-frequency strategy in which the PTA data are first divided into small segments and then SVD is applied to each segment to output the two signal streams. Note that it is usually necessary to allow overlap between successive segments to catch signals occurring in the beginning or end of the segment. The detection of GW bursts of unknown waveforms generally involves searching for any `tracks' or `clusters' of excess power in the time-frequency space \citep[e.g.,][]{tfDetGW99,Wen05EMRI}. So the detection statistic is designed as accumulating signal power in (a `box' of) the time-frequency domain and can be written as:
\begin{equation}
\label{DS-GWb}
\mathcal{P}_{\rm{GWb}}(i,k)= \sum^{l/2}_{a=-l/2}\sum^{m/2}_{b=-m/2} \left(|\mathbf{\tilde{d}}_{1,(i+a)(k+b)}|^{2}+|\mathbf{\tilde{d}}_{2,(i+a)(k+b)}|^{2}\right),
\end{equation}
for the $i$-th segment and $k$-th frequency bin. Here $l$ and $m$ is the length of box in time and frequency respectively. If the data consist of only Gaussian noise, $\mathcal{P}_{\rm{GWb}}$ follows a $\chi^2$ distribution with $4N_{b}$ degrees of freedom (where $N_{b}=l\times m$).

The detection statistics proposed here all follow a noncentral $\chi^2$ distribution with their corresponding degrees of freedom when signals are present in the data. It is convenient to define the expected signal-to-noise ratio ($\rho$) as the noncentrality parameter
\begin{equation}
\label{DS_rho}
\langle\mathcal{P}\rangle= N_{\rm{dof}}+\rho^{2},
\end{equation}
where the brackets $\langle ... \rangle$ denote the ensemble average of the random noise process, $N_{\rm{dof}}$ is the number of degrees of freedom for the `central' distribution of $\mathcal{P}$. We use $\rho$ to quantify the signal strength in our simulations. Note that $\rho^2$ equals the detection statistic calculated for noiseless data. In a frequentist detection framework, we are interested in the FAP of a measured detection statistic $\mathcal{P}_{0}$, i.e., the probability that $\mathcal{P}$ exceeds the measured value for noise-only data. For the aforementioned methods the single-trial FAP is given by $1-\mathrm{CDF}(\mathcal{P}_{0};\chi^{2})$ where $\mathrm{CDF}(\, ;\chi^{2})$ denotes the cumulative distribution function (CDF) for the central $\chi^2$ distribution in question. When a search is performed over unknown source parameters, the total FAP is $1-[\mathrm{CDF}(\mathcal{P}_{\rm{max}};\chi^{2})]^{N_{\rm{trial}}}$ for the maximum detection statistic $\mathcal{P}_{\rm{max}}$ found in the search with $N_{\rm{trial}}$ being the trials factor, which is defined as the number of independent cells in the searched parameter space for a grid-based search.

\subsection{Relation to the `$A_{+}A_{\times}$' method}
\label{sec-ApAc}
In the pulsar timing software package \textsc{TEMPO2} \citep{TEMPO2}, there exists a functionality with which two time series $A_{+,\times}^{{\rm{t2}}}(t)$ can be estimated for a given sky direction. These two time series correspond to two polarizations of the coherent Earth-term timing residuals. Here we call it the `$A_{+}A_{\times}$' method. In this method $A_{+,\times}^{{\rm{t2}}}(t)$ are modelled as time-varying parameters and treated just like any other normal parameters in a timing model so that they can be estimated through a global least-squares fitting routine. As observations for different pulsars are usually unevenly sampled and not at identical times, linear interpolation is used for such a global fit. The `$A_{+}A_{\times}$' method enables one to simultaneously fit for single-source GWs and normal pulsar timing parameters. To avoid the covariance between the global fit for $A_{+,\times}^{{\rm{t2}}}(t)$ and the fit for timing model parameters of individual pulsars, some constraints must be set on the two time series. Currently three kinds of constraints are implemented in \textsc{TEMPO2}, namely, (1) quadratic constraints that correspond to pulsar spin parameters, (2) annual sinusoids for pulsar positions and proper motions, and (3) biannual sinusoids for pulsar parallax. These were first introduced and implemented in \citet{MikeDM2013} when correcting the dispersion measure variations for individual pulsars.

The `$A_{+}A_{\times}$' method was first illustrated for arbitrary GW bursts in fig. 5 of \citet{PPTA13CQG}. It was used in \citet{PPTAcw14} for an all-sky search for monochromatic GWs in the latest PPTA data set. A complete presentation on the `$A_{+}A_{\times}$' method and its applications to single-source GW detection will be provided in a forthcoming paper (Madison et al., in preparation). Here we briefly discuss the relation between the `$A_{+}A_{\times}$' method and the SVD method proposed in this paper. Firstly the principle of both methods is identical since both use the response matrix $\mathbf{F}$ in a similar way and equation (\ref{Aest}) essentially gives the least-squares solution to the two polarization waveforms, i.e., $\hat{\mathbf{A}}$ in equation (\ref{Aest}) is the frequency-domain equivalent of $A_{+,\times}^{{\rm{t2}}}(t)$. The critical difference is in the implementation -- the `$A_{+}A_{\times}$' method works in the time domain whereas the SVD method works in the frequency domain.

There are two advantages of the SVD method over the `$A_{+}A_{\times}$' method:\newline
(1) The SVD method is much faster when searching over the whole sky is required. This is because it works with post-fit residuals after data for each pulsar have been fitted for a timing model and a search over unknown source sky location only involves doing the SVD for the response matrix $\mathbf{F}$, while in the `$A_{+}A_{\times}$' method a global fit is done for each searched sky location (as in the current implementation);\newline
(2) A truncated SVD may be used to improve the detection sensitivity, which applies to the case when $s_{1}\gg s_{2}$, e.g., when the PTA has very low response to one of the two GW polarizations for some sky regions. This is possible especially for current PTAs whose sensitivities are dominated by a few best-timed pulsars.\newline
The `$A_{+}A_{\times}$' method has been fully tested and implemented in real data for continuous GWs in \citet{PPTAcw14}. We will demonstrate the effectiveness of the SVD method with some examples using idealized data sets in section \ref{Sec-Examp} and then more realistic data sets in section \ref{Sec-Realis}.

\section{Examples using idealized data sets}
\label{Sec-Examp}
For the purpose of illustration of our method, we consider an idealized PTA consisting of 20 PPTA pulsars. The simulated observations are evenly sampled once every two weeks with a time span of 10 yr. All observations contain stationary white Gaussian noise with a rms of 100 ns and equal error bars. The simulated data sets are produced as a combination of realizations of white Gaussian noise and Earth-term timing residuals. When we apply our detection statistics to noise-only data sets, it is confirmed that they follow a $\chi^2$ distribution with their respective numbers of degrees of freedom.

In the following three subsections, we illustrate how our method works in terms of detection, localization and waveform estimation for the three types of signals considered in this work. The analysis is simplified as follows: (1) firstly the detection problem is demonstrated by evaluating the detection statistics at the injected source location; (2) then detection statistics are computed on a uniform grid of sky positions and for a range of frequencies to find the maximum; and (3) finally the frequency-domain waveforms $A_{+}(f)$ and $A_{\times}(f)$ are estimated at the actual source sky location. To show the correctness of the estimation process, a number of noise realizations were performed in order to obtain average estimates of $A_{+,\times}(f)$ which were then compared with the true waveforms. For simplicity, in the relevant figures we only plot the absolute values of $A_{+}(f)$ and $A_{\times}(f)$, which are called spectral signatures. For the detection problem we calculate the single-trial FAP to quantify the detection significance, which is only appropriate when source parameters (such as sky location and frequency) are known as we assume for the following examples. The simulated signals are weak to moderately strong with signal-to-noise ratios ranging from 5 to 10.

\subsection{Monochromatic waves}
\label{sec-ExpMon}
\begin{figure}
\centerline{\includegraphics[width=0.6\textwidth]{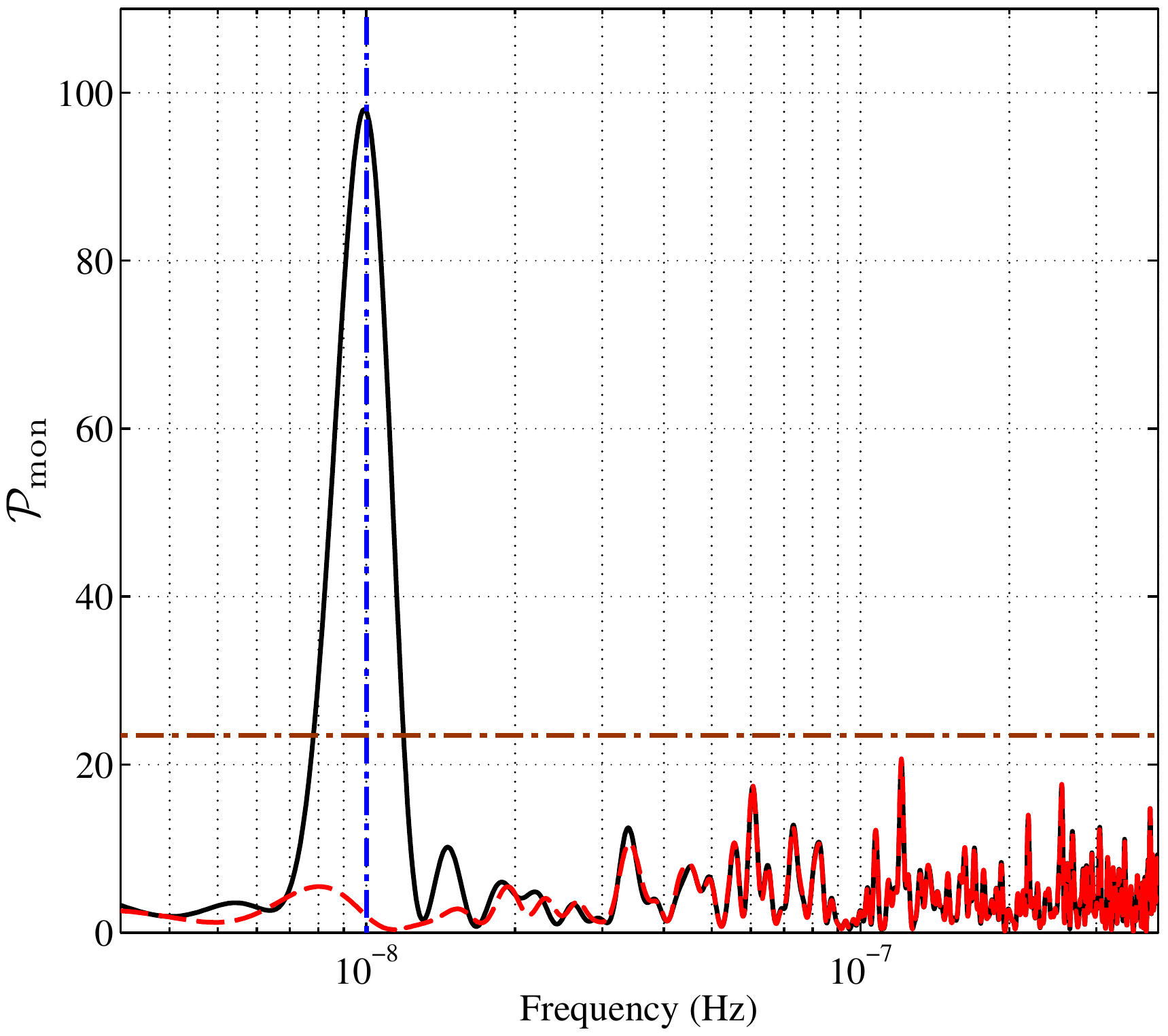}}
\caption{Detection statistics ($\mathcal{P}_{\rm{mon}}$) as a function of frequency for a simulated data set that includes a strong monochromatic signal (solid black) and for the noise-only data set (dashed red). The vertical line marks the injected frequency (10 nHz), while the horizontal line corresponds to a single-trial FAP of $10^{-4}$.}
\label{fig:DSexpMon1}
\end{figure}

\begin{figure}
\centerline{\includegraphics[width=12cm]{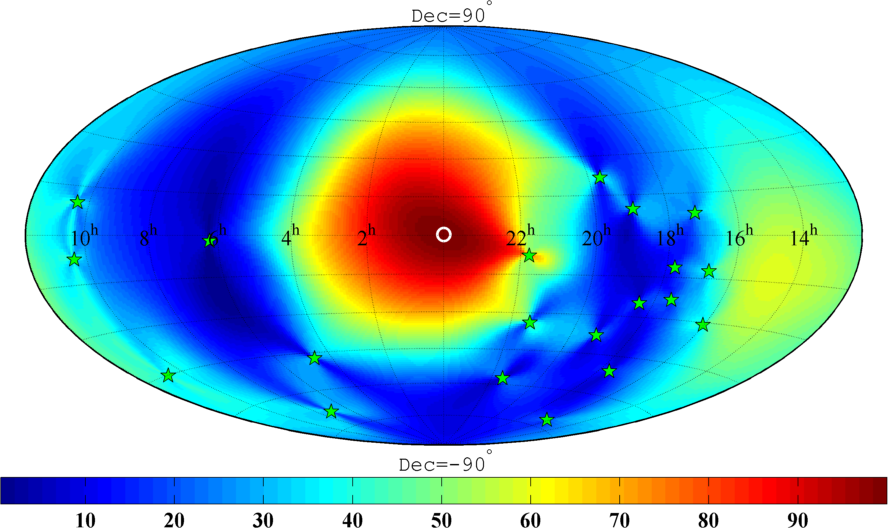}}
\caption{Sky map of detection statistics calculated for a simulated data set that includes a strong monochromatic signal. The signal is simulated at the center of the map and the maximum detection statistic is found at ``$\circ$" (which is the same as the actual source location). Sky locations of the 20 PPTA pulsars are marked with ``$\star$".}
\label{fig:SkyMon1}
\end{figure}

For monochromatic waves, the simulated signal is characterized with the following parameters: $h_{0}=1.16\times 10^{-15}$, $f=10$ nHz, $\cos\iota=1$, $\psi=0$, $\phi_{0}=0$, $(\alpha,\,\delta)=(0,\,0)$. The expected signal-to-noise ratio is $\rho=10$. Fig. \ref{fig:DSexpMon1} shows the detection statistics ($\mathcal{P}_{\rm{mon}}$) as a function of frequency evaluated at the injected sky location. The maximum statistic, which gives extremely strong evidence of a detection, is found at a frequency of 9.94 nHz. To localize this source, we calculate $\mathcal{P}_{\rm{mon}}$ for the same range of frequencies and for a grid of sky directions. At each sky direction, the maximum statistic over frequencies is recorded. Fig. \ref{fig:SkyMon1} shows that the source is successfully localized to where the signal is generated.

\begin{figure}
\centerline{\includegraphics[width=0.6\textwidth]{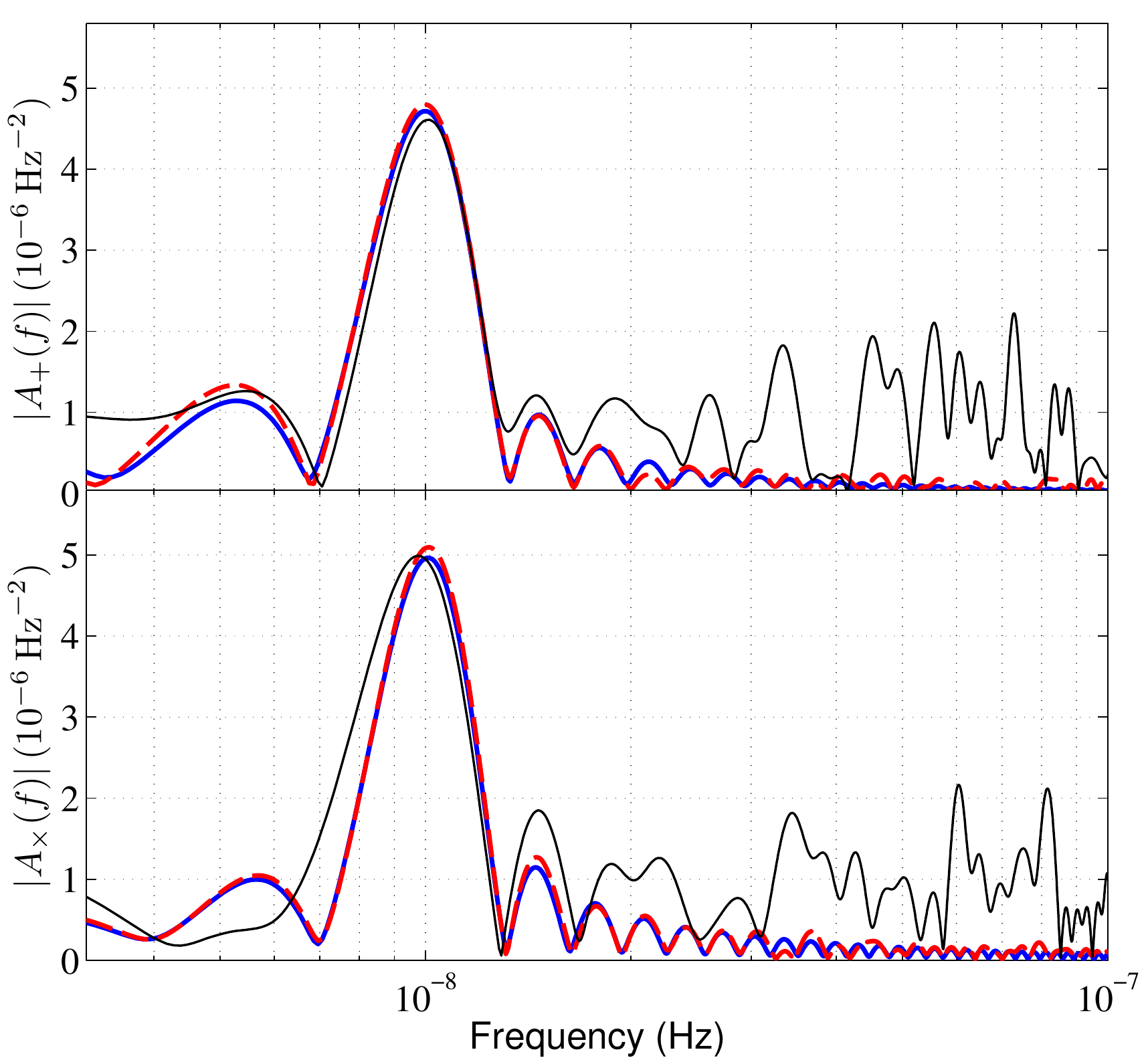}}
\caption{Estimated spectral signatures (thin solid black) for the same data set (white Gaussian noise $+$ a monochromatic GW) used in Figs \ref{fig:DSexpMon1} and \ref{fig:SkyMon1}, along with true spectra (solid blue) and average estimates taken over 100 noise realizations (dashed red).}
\label{fig:RecWaveMon1}
\end{figure}

Having already detected and localized the source, we use equations (\ref{Aest}--\ref{Avar}) to infer the GW waveforms. Fig. \ref{fig:RecWaveMon1} shows the estimated spectral signatures along with the true spectra, indicating a very good estimation as one would expect since the injected signal is very strong. To check the correctness of our method, we overplot in Fig. \ref{fig:RecWaveMon1} the average estimates taken over 100 noise realizations -- they are almost identical to the true spectra.


\subsection{Eccentric binaries}
\begin{figure}
\centerline{\includegraphics[width=0.6\textwidth]{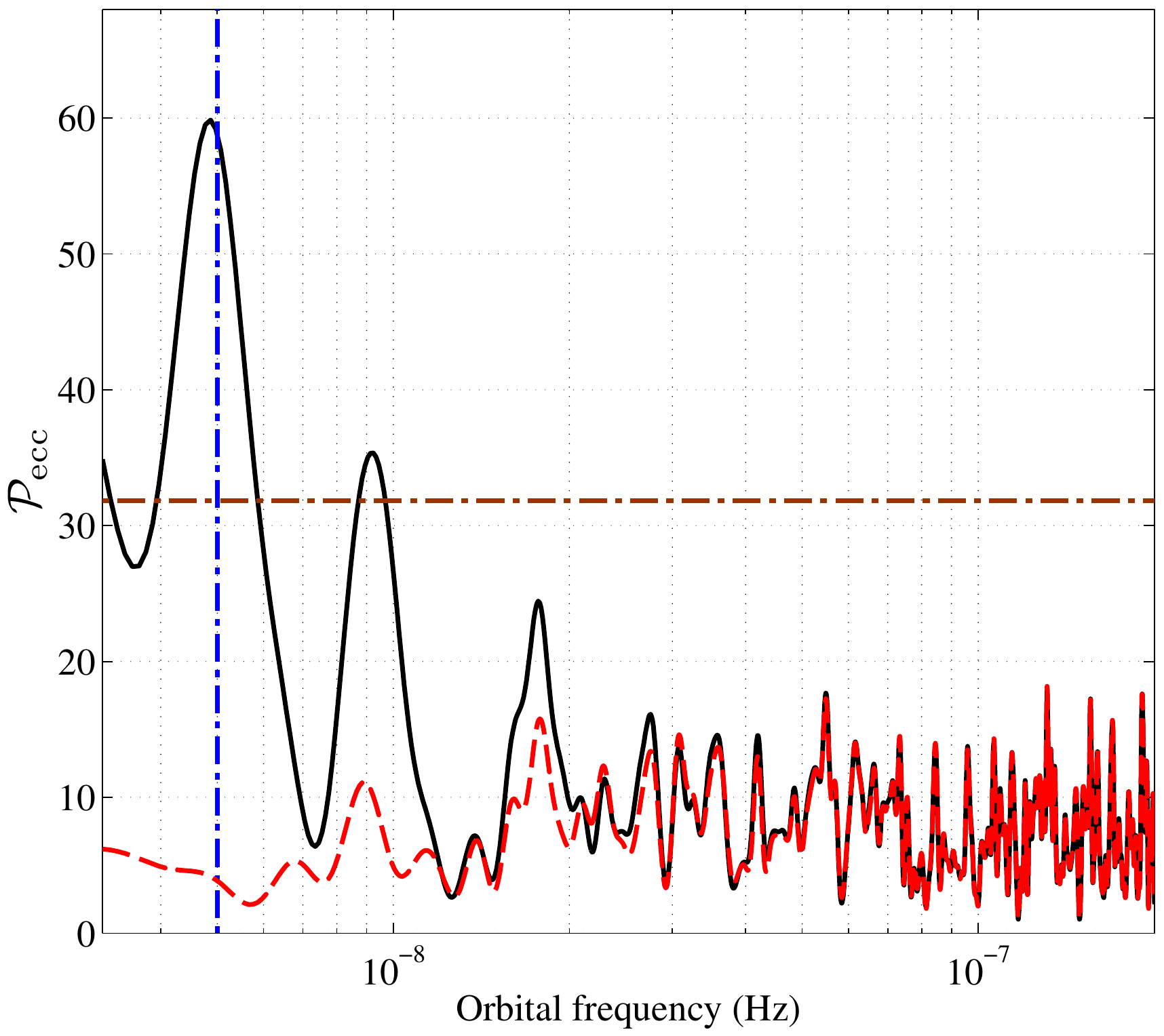}}
\caption{Detection statistics ($\mathcal{P}_{\rm{ecc}}$) as a function of orbital frequency for a simulated data set that includes a moderately strong signal (solid black) and for the noise-only data set (dashed red). The signal is produced by an eccentric SMBBH with an eccentricity of 0.5. Detection statistics are calculated as a harmonic summation up to the second harmonic. The vertical line marks the injected orbital frequency (5 nHz), while the horizontal line corresponds to a single-trial FAP of $10^{-4}$.}
\label{fig:DSexpEcc1}
\end{figure}

\begin{figure}
\centerline{\includegraphics[width=12cm]{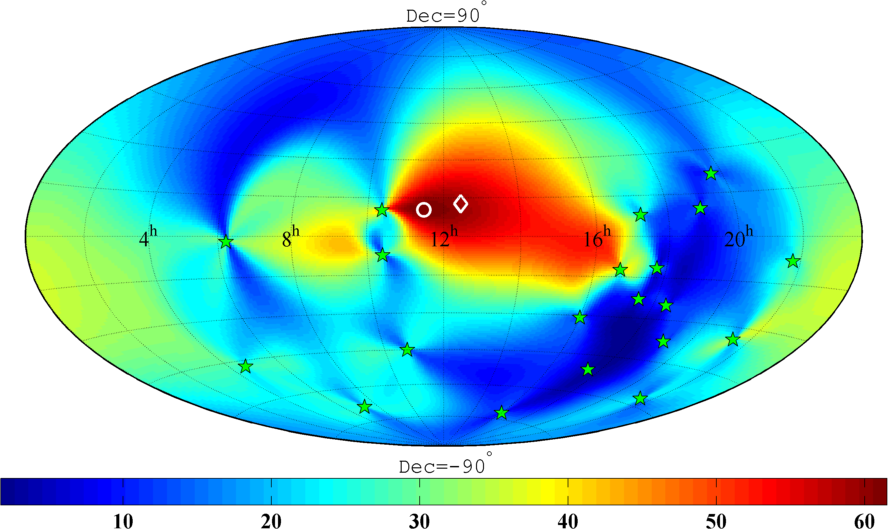}}
\caption{Sky map of detection statistics calculated for a simulated data set that includes a simulated signal produced by an eccentric SMBBH located in the Virgo cluster (marked by a white diamond). The maximum statistic is found at ``$\circ$". Sky locations of the 20 PPTA pulsars are marked with ``$\star$".}
\label{fig:SkyEcc1}
\end{figure}

The simulated source for this example is an eccentric SMBBH located in the Virgo cluster as described by the following parameters: $m_{1}=m_{2}=2\times 10^{8} M_{\odot}$, $d_{L}=16.5$ Mpc, $\cos\iota=1$, $(\alpha,\,\delta)=(3.2594,\,0.2219)$, an orbital frequency of 5 nHz, an initial orbital phase of 1 and an eccentricity of 0.5. The signal is simulated using equations give in section 2 of \citet{JenetWen04}. In order to `detect' this simulated signal, we first apply our analysis at the injected sky location and experiment on the number of harmonics that should be considered. It turns out that a harmonic summation up to the second harmonic gives the lowest FAP. We will further discuss in section \ref{Sec-SenSMBBH} the number of harmonics that should be included for binaries with different eccentricities. Fig. \ref{fig:DSexpEcc1} shows the detection statistics ($\mathcal{P}_{\rm{ecc}}$) as a function of orbital frequency. The maximum statistic 59.85, corresponding to a FAP of $5\times10^{-10}$, is found at 4.87 nHz. The expected signal-to-noise ratio for this injection is $\rho=7$ when we perform a harmonic summing up to the second harmonic. Then we show in Fig. \ref{fig:SkyEcc1} the detection statistics calculated for the whole sky (after maximizing over orbital frequencies for each sky direction). The maximum statistic is found at a grid point close to the injected sky location.

\begin{figure}
\centerline{\includegraphics[width=0.6\textwidth]{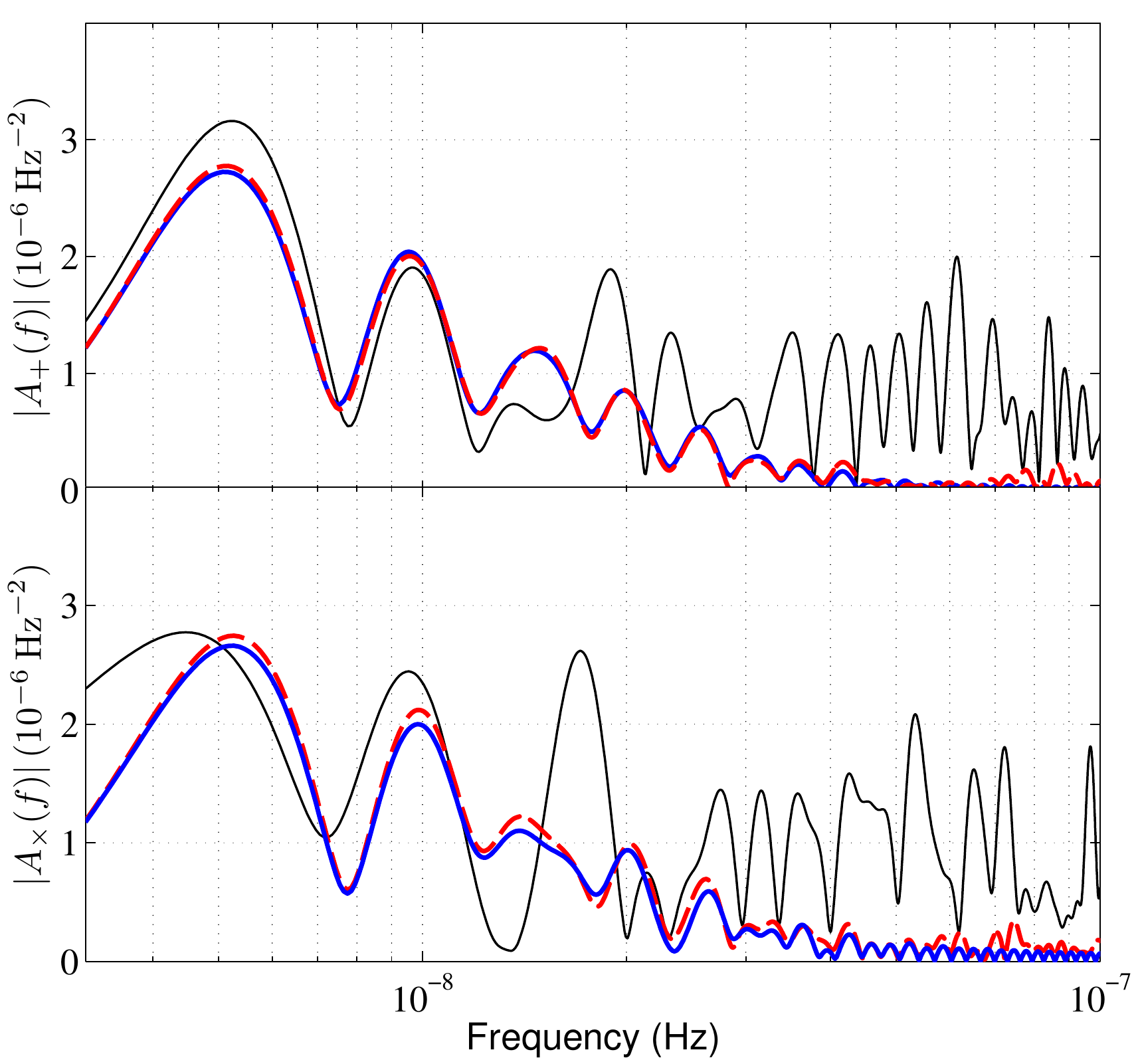}}
\caption{Estimated spectral signatures (thin solid black) for the same data set (white Gaussian noise $+$ a signal expected from an eccentric SMBBH) used in Figs \ref{fig:DSexpEcc1} and \ref{fig:SkyEcc1}, along with true spectra (solid blue) and average estimates taken over 100 noise realizations (dashed red).}
\label{fig:RecWaveEcc1}
\end{figure}

Fig. \ref{fig:RecWaveEcc1} shows the estimated spectral signatures assuming that we know the actual source location. Since the detection statistic varies only slightly within an area of tens of square degrees (as shown in Fig. \ref{fig:SkyEcc1}), similar results should be obtained if we apply the spectral estimation analysis in the sky location where the maximum statistic is found. It is shown that reasonably good estimates are obtained only for the first two harmonics.

\subsection{GW bursts}
\begin{figure}
\centerline{\includegraphics[width=0.6\textwidth]{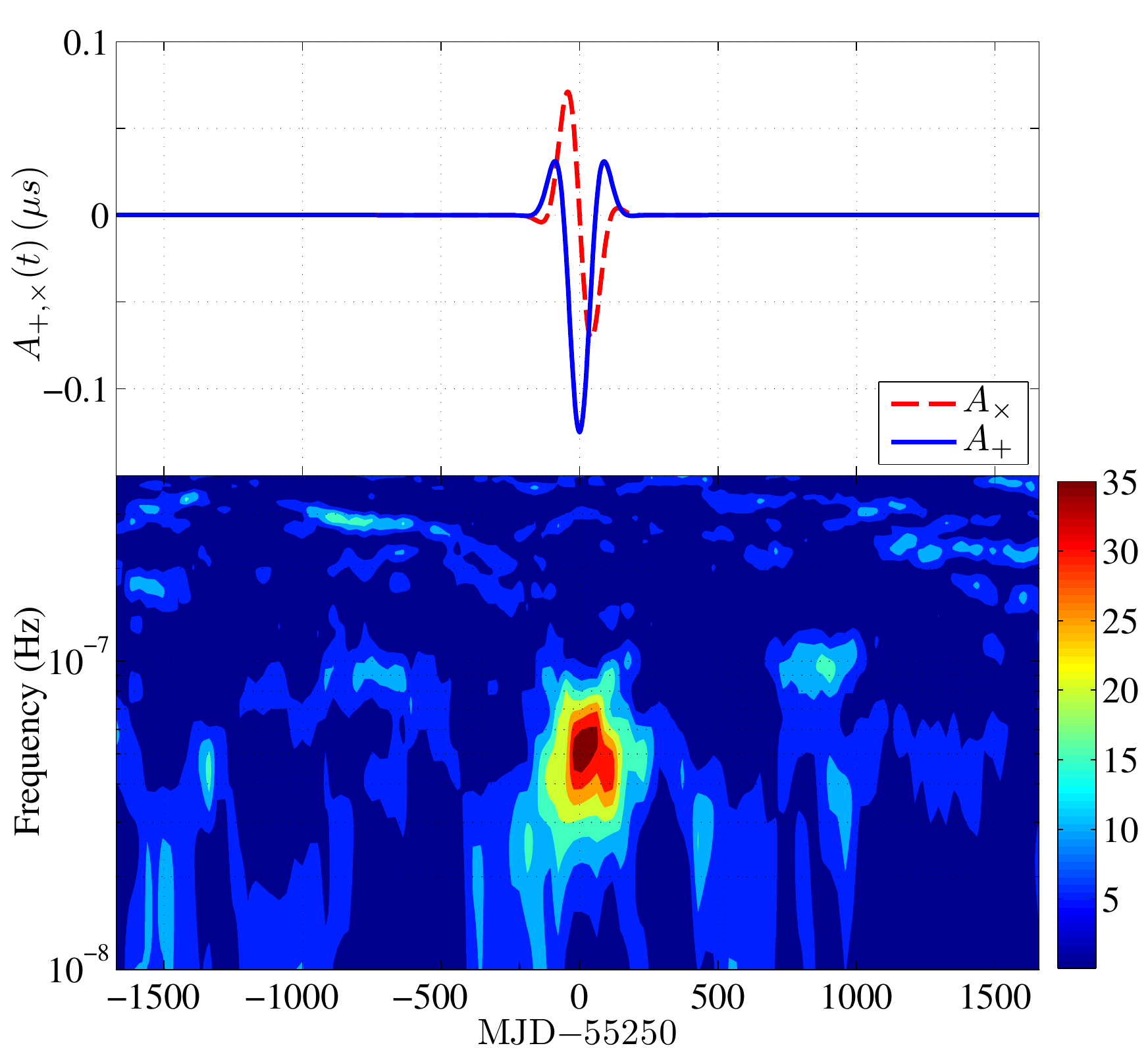}}
\caption{Upper panel: simulated timing signals from a GW burst. Lower panel: detection statistics $\mathcal{P}_{\rm{GWb}}$ (defined in equation \ref{DS-GWb}) calculated in the time-frequency domain.}
\label{fig:DStfGWb1}
\end{figure}

In this example the simulated GW burst signal is described by the following parameters: $A=100$ ns, $\tau=60$ days, $\cos\iota=0.5$, $f_0$=50 nHz, $\psi=1.57$, $\phi_{0}=0$, $(\alpha,\,\delta)=(0.9512,\,-0.6187)$, i.e., originating from the Fornax cluster. The signal occurs in the middle of our observations (MJD 55250). Waveforms of $A_{+,\times}(t)$ are shown in the upper panel of Fig. \ref{fig:DStfGWb1}. For this burst source the amplitudes of timing residuals are $\lesssim 100$ ns, which means the signal should not be visibly apparent in individual pulsar data set.

To dig out this burst signal from noisy data, we divide the 10-yr data set into segments of length of 300 days and calculate the detection statistic given by equation \ref{DS-GWb} in the time-frequency domain. The segment length is (roughly) chosen based on the knowledge of $f_0$, i.e., having $\gtrsim$ one cycle per segment. In practice since the time-frequency analysis for PTA data should not be limited by computational power, many trials of segment length can be performed to search for signals of different durations. The lower panel of Fig. \ref{fig:DStfGWb1} shows results of the time-frequency analysis assuming known source sky location. The most significant statistic 38.36, corresponding to a FAP of $9\times 10^{-8}$, is found at MJD 55257 and a frequency of 51.67 nHz. Therefore our analysis has clearly detected the injected signal and correctly identified its occurrence time and central frequency. For this example, the signal is very `isolated' in the time-frequency space, so it is sufficient to only look at the maximum statistic in the time-frequency map (in this case, the signal-to-noise ratio is $\rho=5$).

\begin{figure}
\centerline{\includegraphics[width=12cm]{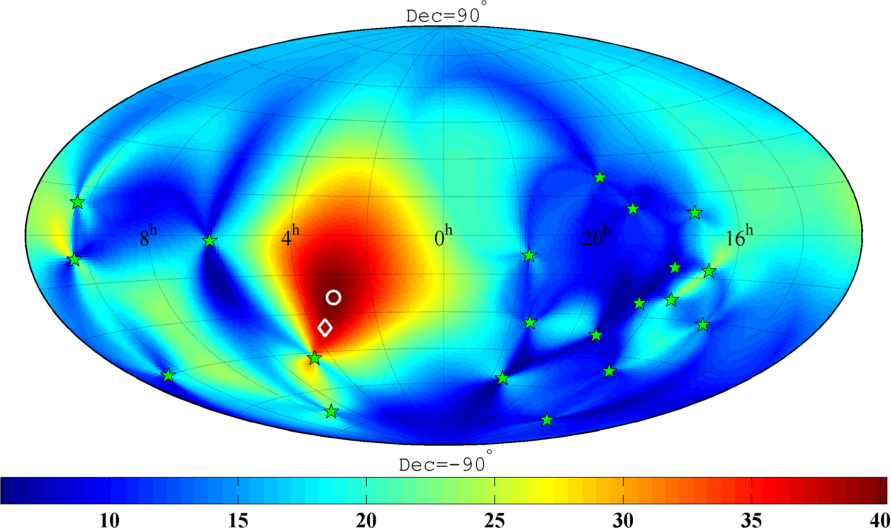}}
\caption{Sky map of detection statistics calculated for a simulated data set that includes a simulated GW burst originating from the Fornax cluster (marked by a white diamond). The maximum statistic is found at ``$\circ$". Sky locations of the 20 PPTA pulsars are marked with ``$\star$".}
\label{fig:SkyGWb1}
\end{figure}

\begin{figure}
\centerline{\includegraphics[width=0.6\textwidth]{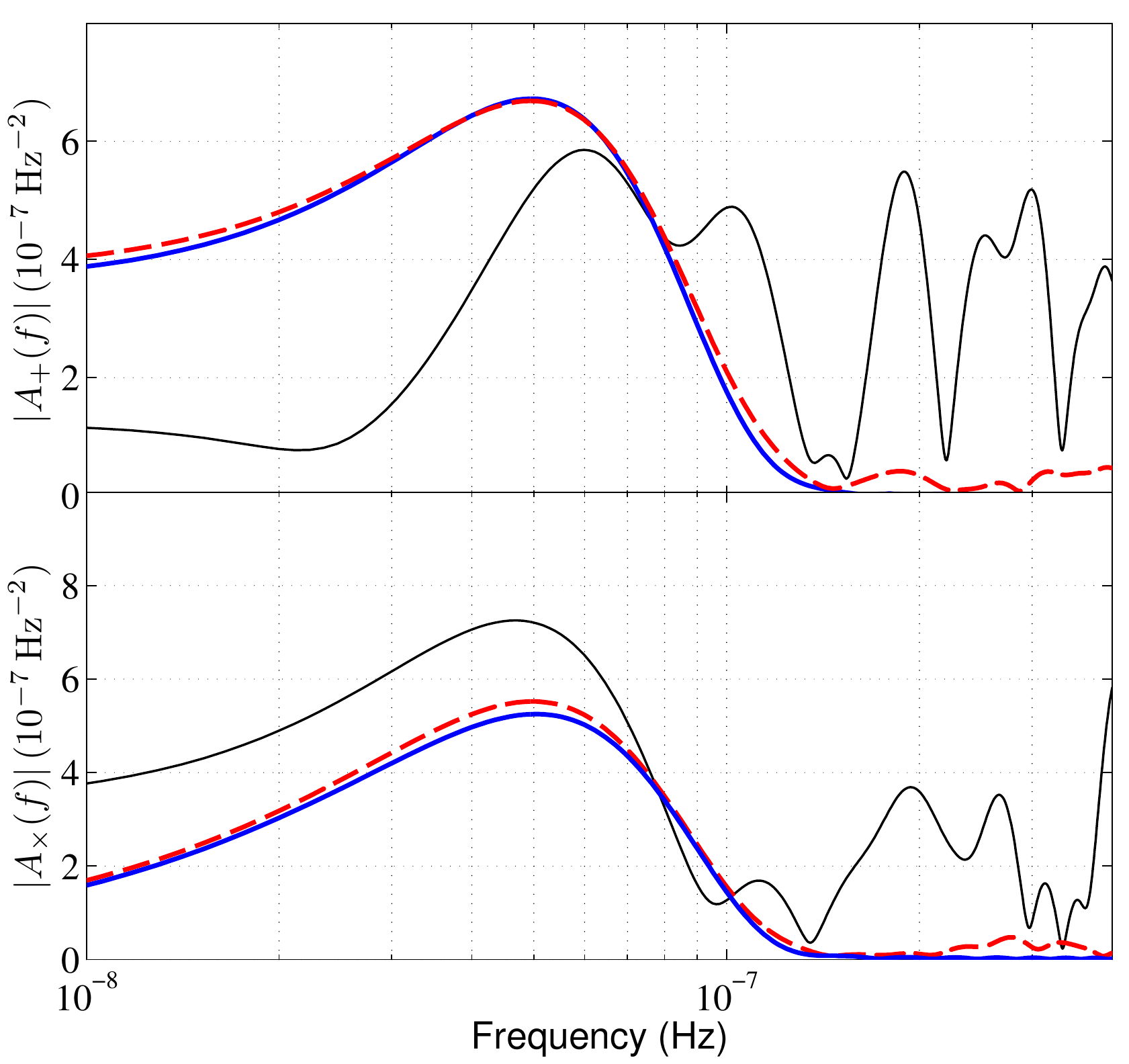}}
\caption{Estimated spectral signatures (thin solid black) for the same data set (white Gaussian noise $+$ a GW burst) used in Figs \ref{fig:DStfGWb1} and \ref{fig:SkyGWb1}, along with true spectra (solid blue) and average estimates taken over 100 noise realizations (dashed red).}
\label{fig:RecWaveGWb1}
\end{figure}

Fig. \ref{fig:SkyGWb1} shows the detection statistics evaluated at the whole sky (after maximizing over frequencies for each sky direction). The maximum statistic is found at a grid point close to the injected source location. Fig. \ref{fig:RecWaveGWb1} shows the inferred spectral signatures compared against the true spectra. As the injected signal is relatively weak for this example, the spectrum is not recovered as well as the previous two examples. For both plots only the central segment that contains the majority of signal power (as illustrated in Fig. \ref{fig:DStfGWb1}) was used.

\section{More realistic data sets}
\label{Sec-Realis}
Examples given in the previous section have assumed idealized PTA observations that are evenly sampled\footnote{Simultaneous to the estimation and correction of the dispersion measure variations for multiple-band pulsar timing data \citep{MikeDM2013}, one obtains estimation of the common-mode signal that is independent of radio wavelengths. These common-mode data sets are evenly sampled and can be searched for GWs.} and contain only stationary white Gaussian noise with equal error bars. Realistic features typical to real PTA data include: (1) observations are irregularly sampled and have varying ToA error bars. It is also common that the data span varies significantly among different pulsars; and (2) low-frequency (`red') timing noise may be present for some pulsars. To simulate realistic data sets, we make use of the actual data spans, sampling and error bars of the PPTA 6-yr Data Release 1 (DR1) data set that was published in \citet{PPTA2013} and used in \citet{PPTAcw14} for an all-sky search for continuous GWs. The PPTA DR1 data set is publicly available online at a permanent link\footnote{\url{http://dx.doi.org/10.4225/08/534CC21379C12}}. As a check, in some simulations an uncorrelated red-noise process with a power-law spectrum is also included. This has no effect on the results since we assume the noise spectrum is known and thus include it explicitly in the noise covariance matrix. In actual analyses of real data, the noise estimation is a very important step and we do not attempt to address it in this work.

\begin{figure}
\begin{center}
        \subfigure[]{%
            \label{fig:PdfMonRl}
            \includegraphics[width=0.6\textwidth]{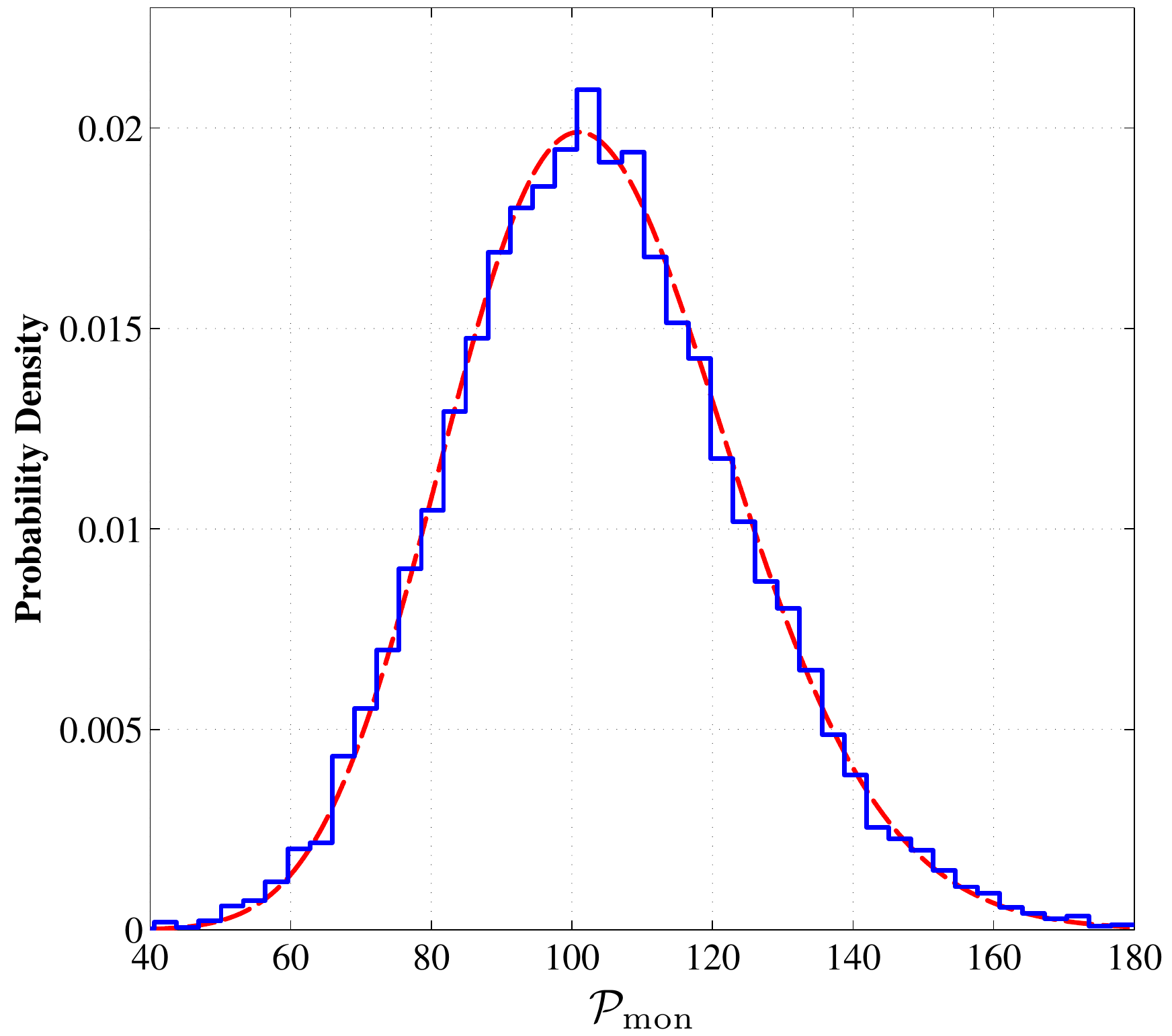}
        }\\
        \subfigure[]{%
            \label{fig:PdfNullMonRl}
            \includegraphics[width=0.6\textwidth]{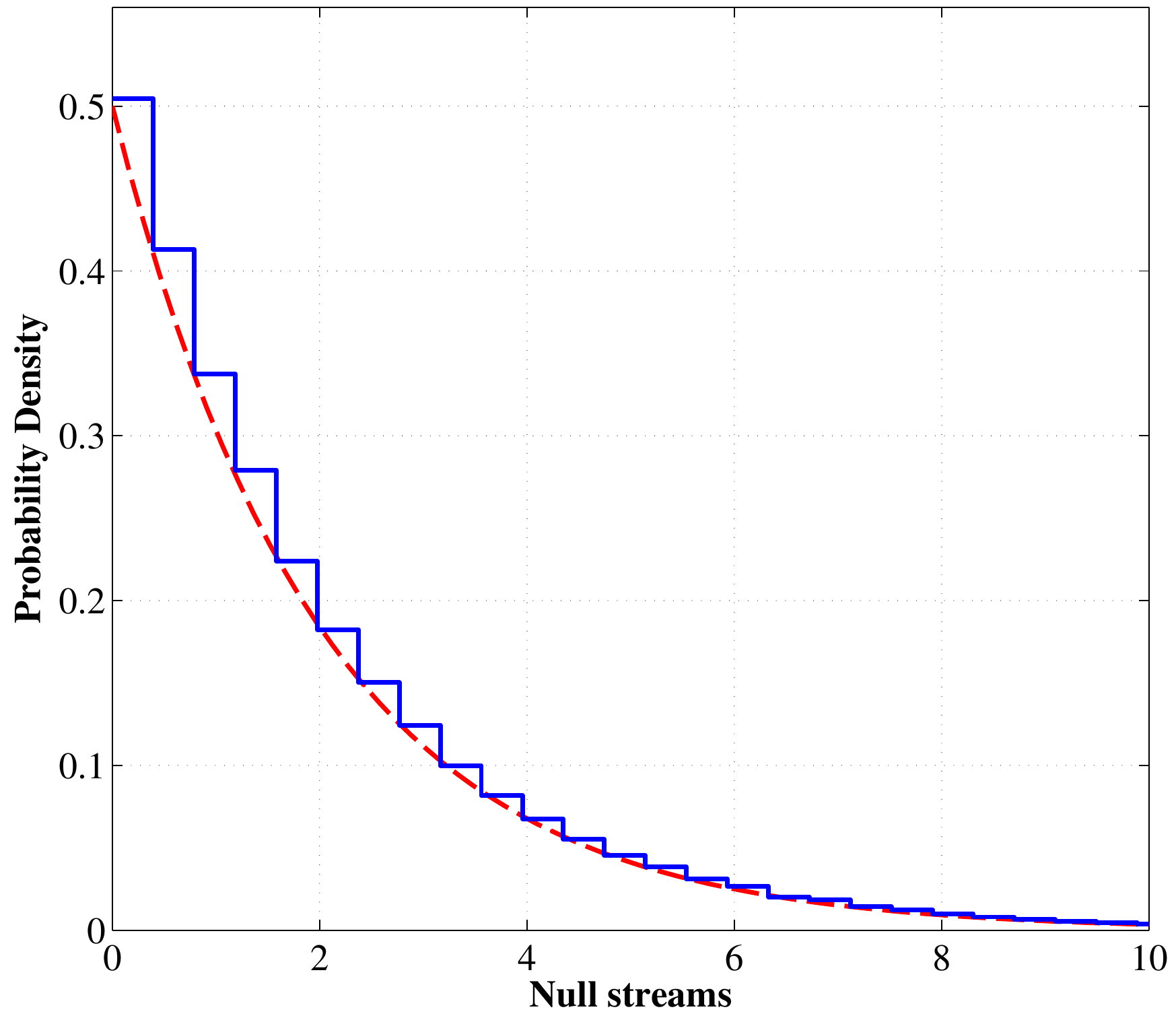}
        }
\end{center}
\caption{(a) Probability distribution of the detection statistic $\mathcal{P}_{\rm{mon}}$ (solid blue) in the presence of a monochromatic signal compared to the expected noncentral $\chi^2$ distribution (dashed red) with 4 degrees of freedom and an noncentrality parameter $\rho^{2}=100$. The simulation was performed with search parameters fixed at the injected values and $10^{4}$ realizations of white Gaussian noise. (b) For the same simulation as panel (a) but instead showing the probability distribution of the squared null streams $|\mathbf{\tilde{d}}_{\rm{null}}|^2$ (solid blue) compared to the expected $\chi^2$ distribution with 2 degrees of freedom (dashed red).}
\label{fig:PdfMonRl1}
\end{figure}

Fig. \ref{fig:PdfMonRl1} shows the distribution of the detection statistics $\mathcal{P}_{\rm{mon}}$ in the presence of signal along with the distribution of corresponding squared null streams ($|\mathbf{\tilde{d}}_{\rm{null}}|^2$) as defined in equation (\ref{new-data1}). The injected signal is characteristic of a circular SMBBH described by the following parameters: $h_{0}=1\times 10^{-14}$, $f=20$ nHz, $\cos\iota=0.6172$, $\psi=4.1991$, $\phi_{0}=1.9756$, $(\alpha,\,\delta)=(0,\,0)$. The expected signal-to-noise ratio is $\rho=10$, which is identical to the example shown in section \ref{sec-ExpMon}. To obtain the distribution of the detection statistic and null streams, we perform $10^{4}$ realizations of white Gaussian noise and for each noise realization keep the search parameters (i.e., source frequency and sky direction) fixed at their injected values. We can see that both match the expected distributions perfectly well. As mentioned in the previous section, null streams can be used as a consistency check in the case of a detected GW signal -- if the detected signal is due to a GW, null streams should follow the expected noise distribution while otherwise false alarms caused by any noise processes are very unlikely to exhibit such a property.

\begin{figure}
\centerline{\includegraphics[width=0.6\textwidth]{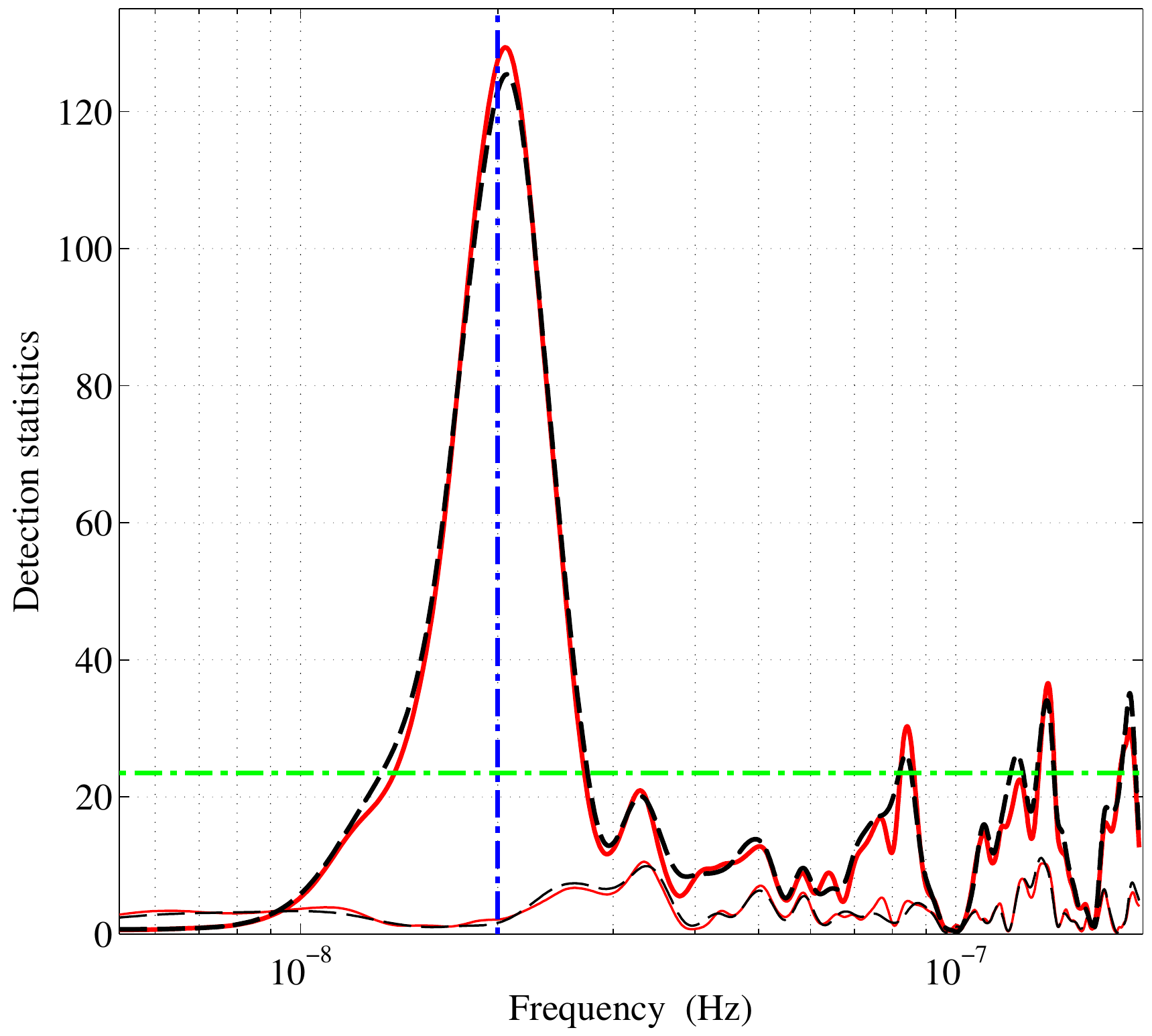}}
\caption{Detection statistics as a function of frequency for realistic simulated data sets in the absence (thin curves) and presence (thick curves) of a strong monochromatic signal. Here we compare two statistics -- $\mathcal{P}_{\rm{mon}}$ (solid red) proposed in this paper and the $\mathcal{F}_{e}$-statistic (black dash). All statistics are evaluated at the injected source sky direction. The vertical line marks the injected frequency (20 nHz), while the horizontal line corresponds to a single-trial FAP of $10^{-4}$.}
\label{fig:DSexpMonRl1}
\end{figure}

Fig. \ref{fig:DSexpMonRl1} shows the detection statistics as a function of frequency calculated for realistic simulated data sets in the absence or presence of the same monochromatic signal as in Fig. \ref{fig:PdfMonRl1}. We have considered two statistics -- $\mathcal{P}_{\rm{mon}}$ proposed in this paper and the $\mathcal{F}_{e}$-statistic that was first proposed in \citet{Babak2012} and later strengthened in \citet{Ellis2012}, both giving nearly identical results. In the signal-present case, the maximum value (129.4) of $\mathcal{P}_{\rm{mon}}$ is found at 20.5 nHz, while the largest $\mathcal{F}_{e}$-statistic (125.4) is measured at 20.6 nHz. It is also obvious that at some high frequencies there is a spectral leakage problem that applies to both methods. This is not surprising given the similar principle of the two statistics: (1) the $\mathcal{F}_{e}$-statistic works fully in the time domain, but it also involves a process of maximum-likelihood estimation of the fourier components in individual pulsar data set; (2) in the derivation of the $\mathcal{F}_{e}$-statistic, the four (time-dependent) basis functions of the monochromatic timing residuals are essentially equivalent to the two orthogonal sine-cosine pairs of $A_{+,\times}(t)$.

\begin{figure}
\centerline{\includegraphics[width=12cm]{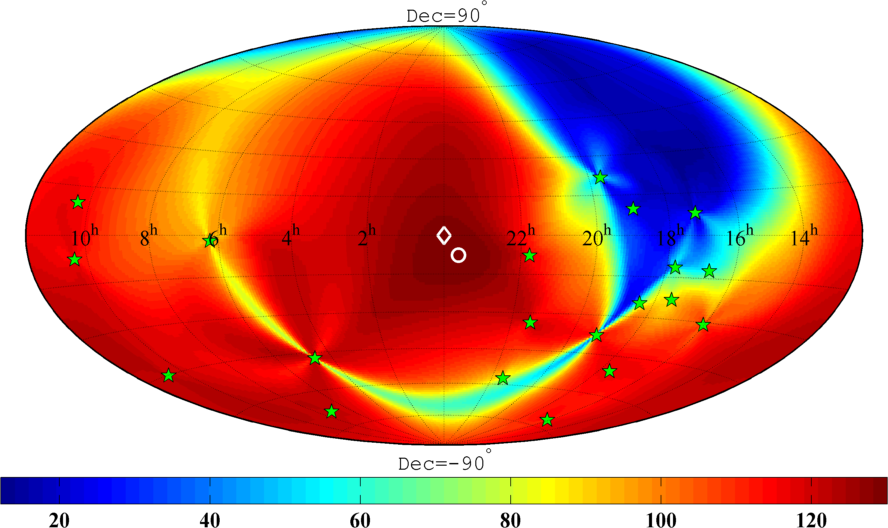}}
\caption{Sky map of detection statistics calculated for a simulated realistic data set that includes a strong monochromatic signal. The source is simulated at the center of the map (indicated by a white diamond) and the maximum detection statistic is found at ``$\circ$". Sky locations of the 20 PPTA pulsars are marked with ``$\star$".}
\label{fig:SkyMonRl1}
\end{figure}

Fig. \ref{fig:SkyMonRl1} shows the results of an all-sky search using the SVD method for the same signal-present data set as used in Fig. \ref{fig:DSexpMonRl1}. The maximum detection statistic 130.1 is found at $(\alpha,\,\delta)=(6.185,\,-0.133)$ with a frequency of 20.5 nHz. The source is localized to a direction close to the injected location. However, compared with Fig. \ref{fig:SkyMon1}, we can see that the angular resolution for realistic PTAs is much worse because the sensitivity is dominated by a few good `timers' in the array. It is worth pointing out that the SVD method is also advantageous in terms of computational efficiency over that of the $\mathcal{F}_{e}$-statistic, since for the latter the amount of computation is proportional to the total number of sky locations being searched.

Similar results were obtained if the simulated data sets used in Figs \ref{fig:PdfMonRl1}--\ref{fig:SkyMonRl1} have gone through the TEMPO2 fitting process for a full timing model for each pulsar. In fact, it has been shown that the fitting process can be approximated by multiplying the timing residuals by a data-independent and non-invertible linear operator matrix \citep[e.g.,][]{DemorestThesis07}. This matrix can be explicitly included in the noise covariance matrix and does not pose a problem for our analysis method. The timing model fit is known to absorb some GW power at low frequencies (close to $1/T_{\rm{span}}$ with $T_{\rm{span}}$ being the data span, because of the fit for pulsar spin period and its first time derivative), and in some narrow bands centred around 1 $\rm{yr}^{-1}$ (the fit for pulsar positions and proper motions), 2 $\rm{yr}^{-1}$ (the fit for pulsar parallax) and high frequencies that correspond to binary periods for pulsars in binaries \citep[e.g., fig. 1 in][]{Cutler14}. These effects are similar to those caused by applying constraints on the $A_{+,\times}^{{\rm{t2}}}(t)$ time series as discussed in section \ref{sec-ApAc}.

\section{Sensitivities to eccentric SMBBHs}
\label{Sec-SenSMBBH}
In this section we are interested in current PTA sensitivities to eccentric binaries. Given that previous published PTA searches for GWs from SMBBHs have assumed zero eccentricity, we attempt to answer the following question -- when is a monochromatic search sufficient to detect eccentric binaries? For this purpose, we use simulated data sets that have the same sampling and error bars as the PPTA DR1 data set, and consider two detection statistics $\mathcal{P}_{\rm{mon}}$ and $\mathcal{P}_{\rm{ecc}}$. The sensitivities are parameterized by the luminosity distance within which $95\%$ of sources are detectable at a fixed FAP of $10^{-4}$.

To facilitate our calculations, we simulate signals due to eccentric binaries drawn from uniform distribution in $\cos\delta$ and $\alpha$ and evaluate both $\mathcal{P}_{\rm{mon}}$ and $\mathcal{P}_{\rm{ecc}}$ for noiseless data at the injected sky location. For the demonstration here, we consider sources with a chirp mass of $10^{9} M_{\odot}$ and an orbital frequency of 10 nHz with other parameters such as $\cos\iota$, initial orbital phase and polarization angle randomized. Since here the data consist of signals only, the detection statistic equals the squared signal-to-noise ratio $\rho^2$ and scales inversely with $d_L^2$. We use this scaling relation to find the value of $d_L$ that corresponds to the given detection threshold. We perform $10^4$ Monte Carlo simulations and choose the $95\%$ quantile as a point in the sensitivity plot.

\begin{figure}
\centerline{\includegraphics[width=0.6\textwidth]{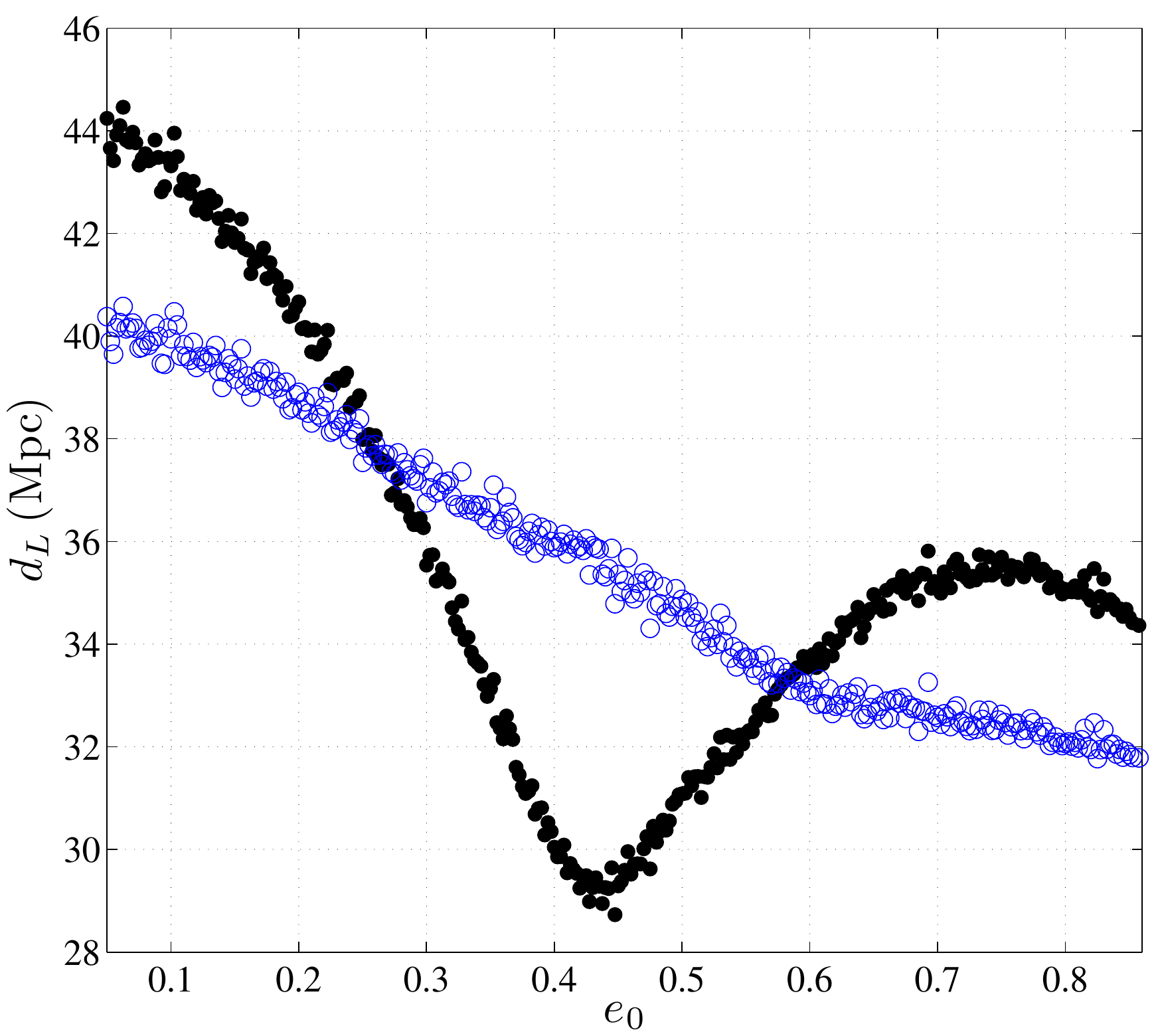}}
\caption{Luminosity distance ($d_L$), within which $95\%$ of eccentric SMBBHs with a chirp mass of $10^{9} M_{\odot}$ and an orbital frequency of 10 nHz could be detectable with the current PTA sensitivity, as a function of orbital eccentricity ($e_0$). Two detection strategies are considered: a monochromatic search (black dots) and a harmonic summation technique (blue open circles). Note that the sensitive distance scales as $M_{c}^{5/3}$.}
\label{fig:SenEcc_Mon1}
\end{figure}

Fig. \ref{fig:SenEcc_Mon1} shows the sensitivity to GWs from eccentric SMBBHs using two detection strategies -- a monochromatic search and a harmonic summing technique. It is clearly demonstrated that for high and low eccentricities a monochromatic search is better than the harmonic-summing search, while the latter is more sensitive in the moderate regime ($0.3\lesssim e_{0}\lesssim 0.55$). This is in good agreement with the fact that for high and low eccentricities the GW emission is dominated by the orbital frequency and its second harmonic respectively. In the high- and low-eccentricity regimes, adding incoherently power from secondary harmonics is not beneficial because when a harmonic summing technique is adopted the detection threshold should be increased for a fixed FAP. We also find that the inclusion up to the third harmonic is sufficient for all possible eccentricities in this example.

\section{Summary and future work}
\label{Sec-Conclu}
PTA experiments have steadily improved their sensitivities over the past few years and produced very stringent constraints on the evolution of ensembles of SMBBHs. While the traditional focus was a stochastic background from SMBBHs throughout the Universe, possibilities of detecting and studying single-source GWs using PTAs have been explored in depth in recent years. Although it may be more challenging, detections of individual sources will provide rich information on the properties of the GW emitters such as the orbital eccentricities, masses and even spins of SMBBHs. This prospect becomes increasingly important as the Chinese
500-meter aperture spherical telescope (FAST) is expected to come online in 2016 \citep{FAST11,GeorgeFAST14} and the
planned Square Kilometer Array (SKA) will be operational within a decade \citep{Lazio13CQG}. Both FAST and SKA will provide a major step forward in terms of single-source GW detection with PTAs.

Based on the coherent network analysis method used in ground-based interferometers, we proposed a method that is capable of doing detection, localization and waveform estimation for various types of single-source GWs. We demonstrated the effectiveness of this method with proof-of-principle examples for GWs from SMBBHs in circular or eccentric orbits and GW bursts. We also demonstrated the implementation of this technique using realistic data sets that include effects such as uneven sampling and varying data spans and error bars. The new method is found to have the following features: (1) it is fast to run especially for all-sky blind searches; (2) it performs as well as published time-domain methods for realistic data sets; and (3) null streams can be constructed as a consistency check in the case of detected GW signals. Finally, we presented sensitivities to eccentric SMBBHs and found that (1) a monochromatic search that is designed for circular binaries can efficiently detect SMBBHs with both high and low eccentricities; and (2) a harmonic summing technique provides better detection sensitivities for moderate eccentricities.

Our future work will be:\newline
(1) to develop a fully functional data analysis pipeline that can be tested in future IPTA mock data challenges\footnote{\url{http://www.ipta4gw.org/}} and applied to real data;\newline
(2) to make comprehensive comparisons (in terms of detection sensitivity, parameter estimation and speed) between the method proposed in this paper and published methods such as the `$A_{+}A_{\times}$' method and the $\mathcal{F}_{e}$-statistic. Again this could be done along with a future IPTA data challenge;\newline
(3) to include pulsar terms in the detection framework for continuous GWs and investigate how the information of pulsar terms can be exploited to improve the PTA angular resolution.

\section*{Acknowledgments}
We thank the referee for useful comments. This work was supported by iVEC through the use of advanced computing resources located at iVEC@UWA. We are grateful to Bill Coles and Yan Wang for useful discussions. X-JZ acknowledges the support of an University Postgraduate Award at UWA. LW acknowledges funding support from the Australian Research Council. GH is supported by an Australian Research Council Future Fellowship. J-BW is supported by West Light Foundation of CAS XBBS201322 and NSFC project No.11403086.

\bibliographystyle{mn2e}
\bibliography{Ref}
\label{lastpage}
\end{document}